\documentstyle[preprint,aps]{revtex}

\begin{document}
\draft
\preprint{}

\title{Giant Magneto-Resistance in Nd$_{0.7}$Sr$_{0.3}$MnO$_3$
 at Optical Frequencies}
\author{S. G. Kaplan$^{(a)}$\cite{byline}, M. Quijada$^{(a)}$, H. D. Drew$^{(a,b)}$, 
D. B. Tanner$^{(c)}$
G. C. Xiong$^{(b)}$, R. Ramesh$^{(b)}$, C. Kwon$^{(b)}$, 
and T. Venkatesan$^{(b)}$}
\address{$^{(a)}$Laboratory for Physical Sciences, College Park, Maryland 20740
}
\address{$^{(b)}$Center for Superconductivity Research, Department of Physics,
University of Maryland at College Park, Maryland 20742}
\address{$^{(c)}$Department of Physics, University of Florida, Gainesville, 
FL 32611}
\date{\today}
\maketitle

\begin{abstract}
	The optical properties of Nd$_{0.7}$Sr$_{0.3}$MnO$_3$ thin films have 
been studied from 5 meV to 25 meV and from 0.25 eV to 3 eV, at temperatures 
from 15 K to 300 K and magnetic fields up to 8.9 T.  A large transfer of 
spectral weight 
from high energy to low energy occurs as the temperature is decreased below 
180 K, where the dc resistivity peaks, or as the magnetic field is increased.  
The optical data are found to be consistent with models that include both the 
double exchange interaction and the dynamic Jahn-Teller effect on the 
Mn$^{3+}$ e$_g$ 
levels.
\end{abstract}


	The recent discovery of colossal magneto-resistance (CMR) in hole-doped 
ferromagnetic manganite materials of the form 
(Ln)$_{1-x}$(A)$_x$MnO$_{3-d}$, where 
Ln is a 
lanthanide and A is an alkaline-earth element, has revived interest in this 
complex magnetic system\cite{kuster1}.  A decrease in resistance of more than 99\% has 
been observed at temperatures of $\approx 150$ K and fields of $\sim 
8$ T in thin films of 
Nd$_{0.7}$Sr$_{0.3}$MnO$_3$\cite{2}.  Strong coupling 
between the electronic and ionic degrees of freedom leads to a static 
Jahn-Teller (J-T) 
distortion of LaMnO$_3$ and a magnetic-field induced structural phase 
transition in La$_{0.825}$Sr$_{0.175}$MnO$_3$\cite{3}.  Also, in a recent optical 
reflectivity study of La$_{0.825}$Sr$_{0.175}$MnO$_3$ 
\cite{4}, a large transfer of spectral weight from high frequencies to low 
frequencies is observed as the sample is cooled from the paramagnetic state 
through the Curie temperature, and explained by optical selection rules within 
a double-exchange picture \cite{5} of the ferromagnetic state.  However, a recent 
theoretical work has pointed out shortcomings of this simple picture and 
suggested the important role of dynamic J-T distortions \cite{6} in the lattice in 
order to fully explain the transport properties of CMR manganites.
	
In this paper we present transmittance and reflectance measurements of 
Nd$_{0.7}$Sr$_{0.3}$MnO$_3$ thin films as a function of both temperature and 
magnetic field.  Our results also show large shifts in spectral weight from 
above 1 eV to lower energy as the temperature is lowered below 180 K, or as 
the magnetic field is increased, demonstrating a broadband change in 
electronic properties at energies several orders of magnitude larger than either 
$\mu_B gH \approx 1 $ meV or k$_B$T.  In contrast to 
La$_{0.825}$Sr$_{0.175}$MnO$_3$, 
the optical properties of Nd$_{0.7}$Sr$_{0.3}$MnO$_3$ show evidence for 
combined double-exchange and dynamic J-T effects.
	
The samples used in this study were Nd$_{0.7}$Sr$_{0.3}$MnO$_3$ thin films 
grown on LaAlO$_3$ and Al$_2$O$_3$ substrates by pulsed laser deposition in an 
N$_2$O atmosphere \cite{2}.  The films are epitaxial as revealed by x-ray 
diffraction, and show 3 MeV He$^+$ ion Rutherford backscattering channeling 
spectra with a minimum yield of 3.8\%, indicating a high degree of 
crystallinity.  For the sample discussed below, grown on an Al$_2$O$_3$ 
substrate, the room temperature dc resistivity measured with a four-probe 
method was 2$\times 10^{-3}~\Omega$-m.  The maximum resistivity, at 180 K, was 
2.4$\times 10^{-2}~\Omega$-m.

Transmittance and reflectance measurements were performed with a combination 
of Fourier transform spectrometers and grating monochromators to cover the 
investigated regions of 5 meV to 25 meV and 0.25 eV to 3 eV.  High magnetic 
field transmittance measurements in the Faraday geometry were made with a 9 T 
superconducting solenoid, using either a lightpipe or optical fiber to gain 
access to the bore of the magnet.  The combined relative uncertainty in the 
absolute transmittance is $\pm 5$\% from 5 meV to 25 meV and $\pm 8$\% 
from 0.25 eV to 
3 eV. (Two standard deviations are used in reporting uncertainties in this 
paper.)  The relative uncertainty in the reflectance spectra is estimated at 
$\pm 6$\%.

Transmittance and reflectance spectra at zero magnetic field from 15 K to 300 K 
for a 140 nm $\pm$10 nm thick Nd$_{0.7}$Sr$_{0.3}$MnO$_3$ film on a 0.5 mm 
thick Al$_2$O$_3$ substrate are shown in Fig.~\ref{fig1}.  The inset in 
Fig.~\ref{fig1}(b) shows 
the low frequency transmittance of this film from 15 K to 240 K, relative to a 
blank Al$_2$O$_3$ substrate. The free-carrier absorption is very weak in this 
sample, which is clear from the strong phonon feature that appears at 21 meV.  
Since the low-frequency transmittance of the film is different from 1 by less 
than 35\%, we can expand the standard sheet conductance formula to linear terms 
in conductivity, $\sigma$, and find that
\begin{equation}
{T_{FS}\over {T_S}} \approx 1 - {2Z_0\sigma_1d\over n+1}\label{one}.
\end{equation}
Here T$_{FS}$ and T$_S$ are the transmittance of the film/substrate and substrate, 
respectively, n = 3.5 is the index of refraction of the substrate, d is the 
film thickness, $\sigma_1$ is the real part of the film conductivity, and Z$_0$ = 377 
$\Omega$ 
per square is the impedance of free space.  Thus, the low-frequency 
transmittance values can be used to directly obtain the low-frequency 
conductivity.

The room temperature conductivity at 5 meV is approximately 10 times larger 
than the measured dc value of this sample, indicative of the granular nature 
of the conductance in this type of thin film material.  The 5 meV conductivity 
has a minimum as a function of temperature at 180 K, showing the same trend as 
the dc conductivity.  Similar behavior has been found for the microwave 
conductivity of these materials \cite{7}.  Also, the conductivity increases as a 
function of frequency, showing no evidence of a Drude-like roll-off.

The conductivity from 0.25 eV to 3 eV is obtained from the transmittance and 
reflectance curves shown in Fig.~\ref{fig1} by numerically inverting the exact 
Fresnel formulas for a thin film on a thick substrate, treating the film 
coherently and the substrate incoherently \cite{8}.  The combined uncertainty in 
the derived conductivity values, which is dominated by the uncertainty in the 
film thickness, is $\pm 20$\%.

The real part of the optical conductivity at zero field derived from the data 
in Fig.~\ref{fig1} is shown in Fig.~\ref{fig2}.  The symbols near the origin show the 
conductivity at 5 meV, while the solid and dashed curves show the conductivity 
derived from the high 
frequency transmittance and reflectance data.  The room temperature curve shows 
an absorption band centered near 1.2 eV, and the beginning of another band 
beyond the high energy limit of the data.  As the temperature is lowered below 
180 K, the conductivity at lower energy increases dramatically, while 
decreasing somewhat at higher energy.  At the same time, the peak in the 
conductivity shifts to lower energy and broadens.

A comparison of the temperature dependence and the magnetic-field dependence 
of the transmittance is shown in Fig.~\ref{fig3}.  Fig.~\ref{fig3}(a) shows the 
ratio of the 
transmittance at a given temperature to that at 180 K, for temperatures below 
180 K, and Fig.~\ref{fig3}(b) the ratio of transmittance at 8.9 T to that at 0 
T over the same temperature range.  A striking similarity can be seen between 
Fig.~\ref{fig3}(a) and Fig.~\ref{fig3}(b), indicating that the spectral changes 
upon decreasing the 
temperature relative to 180 K, or upon increasing the applied field at a given 
temperature, are due to the change in spin alignment.

The peak in the magneto-transmittance effect is at 180 K, the same temperature 
as the peak in the dc and 5 meV resistivities.  The magnetic-field effect 
decreases at both lower and higher temperatures, decreasing to about half the 
peak value by 240 K, and 20\%\ of the peak value by 15 K.  The fact that a 
significant magnetic field dependence persists in this sample down to the 
lowest measured temperature is an indication of magnetic disorder in the film 
(or, possibly, of a canted ferromagnetic ground state) as noted by others 
\cite{7,9,10}.

To make a more quantitative comparison between the magnetic-field and 
temperature dependence of the optical conductivity, we fit the conductivity at 
zero field with a sum of a Drude and Lorentzian terms, and use this fit as a 
basis for fitting the high-field transmittance.  These results are shown in 
Fig.~\ref{fig4}(a) as the fitted change in $\sigma_1$ at 180 K between 8.9 T 
and 0 T, compared to the difference between the measured zero-field 
$\sigma_1$ at 15 K and 180 K, in Fig.~\ref{fig4}(b).  The similarity of the two 
curves reaffirms the conclusion that the changes in conductivity induced by 
lowering the temperature or raising the magnetic field are driven by the spin 
polarization of the system.

It is interesting to compare our observed optical conductivity versus 
temperature, shown in Fig.~\ref{fig2}, with other available optical data on 
rare-earth manganites.  At room temperature, the conductivity in single-crystal 
La$_{0.825}$Sr$_{0.175}$MnO$_3$\cite{4} has a broad peak at 1 eV and falls off 
almost linearly to a small dc value.  In the ferromagnetic state, spectral 
weight is transferred from high frequency to low frequency leading to a nearly 
flat conductivity up to 2.5 eV.  The transfer of oscillator strength from high 
to low energy as the spin polarization is increased has been interpreted as 
the suppression of optical charge transfer transitions between the lower and 
upper exchange-split e$_g$ bands, accompanied by an increased Drude conductivity 
due to an enhanced site-to-site hopping amplitude, within a Hubbard-like 
representation of the double-exchange mechanism.

The scale of the high-frequency conductivity in the two materials is similar.  
However, in our case, we do not observe a Drude-like frequency dependence in 
the low-frequency conductivity at any temperature.  Our data is dominated by a 
peak in the conductivity, which decreases in spectral weight and shifts to higher 
energy as the temperature is raised through the magnetic phase transition.  
Also, the room temperature conductivity at 5 meV in our sample is a factor of 
5 lower than in the single crystal La$_{0.825}$Sr$_{0.175}$MnO$_3$.  Lower 
values of the dc conductivity in the Nd versus La systems have been noted in 
dc transport studies \cite{10} and are believed to indicate an intrinsic difference 
in the conductivity of the two alloys.

The qualitative effect of transfer of oscillator strength from high frequency 
to low frequency as a function of spin polarization appears to be 
characteristic of CMR manganites;  we have found a magneto-optical response 
very similar to that shown in Fig.~\ref{fig3}(b) in a La$_{0.7}$Ba$_{0.3}$MnO$_3$ 
film at 240 K.  This sample has a measured dc resistivity about 2 orders of 
magnitude smaller than the Nd$_{0.7}$Sr$_{0.3}$MnO$_3$ film, and shows almost 
no magneto-optic response at low temperature, consistent with a nearly 
saturated ferromagnet.

In the inset to Fig.~\ref{fig2} we show the temperature dependence of the oscillator 
strength expressed in terms of the effective electron density 
\begin{equation}
N_{eff}(\omega) = {2V_{cell}m\over \pi e^2}\int_0^{\omega} 
\sigma_1(\omega^{\prime})d\omega^{\prime}\label{two}
\end{equation}
integrated to 1.5 eV (where $\Delta \sigma = 0$). V$_{cell}$ is the unit cell 
volume.  It is apparent from both Fig.~\ref{fig2} and Fig.~\ref{fig4} that the 
oscillator strength sum rule is not satisfied over the range of photon energies 
that we have investigated.  As the temperature is lowered, the oscillator 
strength above 1.5 eV decreases, but more slowly than the growth at lower 
frequency (and more slowly than is found in La$_{0.825}$Sr$_{0.175}$MnO$_3$).  
Thus, the magnetic field and temperature dependence must extend to higher 
energies.

The increase of the spectral weight of the peak near 1 eV as the temperature is 
lowered below T$_c$ shows that this feature cannot be due simply to d-d 
transitions between spin-split Mn e$_g$ levels. On the other hand, its strong 
temperature and magnetic-field dependence implies that it must involve the 
e$_g$ 
states in an important way.  Indeed, both the initial and final states must 
have a large e$_g$ component.  If, for example, the absorption peak was due to 
transitions from the O(2p) states to the e$_g$ states, then both O(2p) spins 
would participate, and the transition would not depend on the orientation of 
the Mn core spins.  Moreover, the magnetic field dependence means that the 
transition is sensitive to the relative alignment of two adjacent core spins.  
Thus, we conclude that the main component of the conductivity peak is a charge 
transfer transition between the e$_g$ levels of two adjacent sites.  For the case 
where the final state is on an already occupied site (as in LaMnO$_3$), the 
transition energy would be shifted to well above the investigated region by 
the on-site Coulomb energy U $\approx 4 $ eV.  Therefore, we conclude that it must be 
due to the charge transfer transition from a Mn$^{3+}$ e$_g$  level to the unoccupied 
Mn$^{4+}$ e$_g$  levels on an adjacent site.

Millis, Mueller, and Shraiman \cite{11} suggest that this conductivity peak is to be 
interpreted as the transition from a lower dynamic Jahn-Teller split Mn$^{3+}$
e$_g$  level to the unsplit Mn$^{4+}$ e$_g$ levels of a neighboring site.  In their 
model, which includes both double-exchange and J-T polaron effects, the 
splitting depends on the ratio of the Jahn-Teller self-trapping energy to the 
e$_g$ bandwidth \cite{6,11}.  As the spins align, the double-exchange-driven 
hopping probability increases, leading to an increased e$_g$ bandwidth and thus a 
decrease of the  J-T splitting.  The transition energy is expected to be larger 
than half of the J-T splitting because of the difference between the centers of 
gravity of the e$_g$ levels on 
the Mn$^{4+}$ and Mn$^{3+}$ ions.  Optical conductivity curves calculated 
within this model show shifts in oscillator strength, linewidth, and peak 
position versus temperature which are similar to the data in Fig.~\ref{fig2} 
\cite{11}.

Optical transitions involving O(2p) orbitals are also expected in the spectral 
range from 0.25 eV to 3 eV, as suggested by others [4,9].  Therefore, the 1.2 
eV peak in Fig.~\ref{fig2} could contain contributions from several optical transitions.  
However, as argued above, these transitions would not have the observed 
temperature and magnetic field dependence.  Appropriate measurements and 
systematic doping studies will be required to properly differentiate the 
various contributions to the optical conductivity.
	
It is interesting to reconsider the differences between 
Nd$_{0.7}$Sr$_{0.3}$MnO$_3$ and La$_{0.825}$Sr$_{0.175}$MnO$_3$ within the 
double-exchange/J-T polaron model.  The smaller ion size for Nd (compared with 
La) leads to greater Mn-O-Mn bond angle distortions, so that a stronger J-T 
coupling is expected for Nd$_{0.7}$Sr$_{0.3}$MnO$_3$\cite{12}.  This is expected to lead to a 
larger J-T gap.  Therefore, it appears that the difference in the 
low-temperature behavior of the transport and optical properties of the two 
materials can be understood in terms of a finite J-T gap in 
Nd$_{0.7}$Sr$_{0.3}$MnO$_3$ and a near zero gap in 
La$_{0.825}$Sr$_{0.175}$MnO$_3$.  The larger J-T gap may also be responsible 
for the differing behaviors above 1.5 eV, where the conductivity in 
Nd$_{0.7}$Sr$_{0.3}$MnO$_3$ has much less temperature dependence.  This indicates that 
satisfying the optical oscillator strength sum rule in the Nd alloy may 
require a wider frequency range than in the La alloy.

We would like to thank R. Liu, J. P. Rice, and R. U. Datla for use of their 
optical magnet cryostat system in the early stages of this work.  We would 
also like to thank A. J. Millis for communicating the results of the model 
discussed in the text to us prior to publication.  In addition, we acknowledge 
useful discussions with:  S. M. Bhagat, S. E. Lofland, S. Wu, E-J. Choi, R. L. 
Greene, K. L. Empson, R. S. Decca, and V. Kostur.


\begin{figure}
\caption{Reflectance (a) and transmittance (b) of a 140 nm 
thick Nd$_{0.7}$Sr$_{0.3}$MnO$_3$ film 
on Al$_2$O$_3$ at zero magnetic field for various temperatures.  The inset in 
(b) shows the FIR transmittance; in this case the solid line is the 
transmittance at 240 K rather than 300 K.}
\label{fig1}
\end{figure}

\begin{figure}
\caption{Real part of the optical conductivity, 
$\sigma_1$, versus photon energy derived from the data in 
Fig.$~1$.  
A large transfer of spectral weight from high to low energy is apparent, along 
with a shift in the peak to lower energy as the temperature is decreased.  
Inset:  $N_{eff}$  integrated up to 1.5 eV versus temperature.}
\label{fig2}
\end{figure}

\begin{figure}
\caption{ Transmittance ratios for temperature (a) and 
magnetic field (b) for the 
140 nm  Nd$_{0.7}$Sr$_{0.3}$MnO$_3$ sample.  Frame (a) shows the ratio of the 
transmittance at several temperatures to that at 180 K at zero field, while (b) 
shows the ratio for H = 8.9 T to 0 T at 180 K and below.  In frame (b), the solid 
circle shows the FIR transmittance ratio at 15 K, and the diamond the ratio at 
150 K.}
\label{fig3}
\end{figure}

\begin{figure}
\caption{Comparison of (a) the fitted difference 
$\Delta \sigma_1 = \sigma_1 (180 K, 
8.9 T) - \sigma_1 (180 K, 0 T)$ with (b) the measured difference 
$\Delta \sigma_1 = \sigma_1 (15 K, 0 T) - \sigma_1 (180 K, 0 T)$.  The change 
in spectral weight has a similar shape in both cases, with the large increase 
at low energy implying that the decrease above 1.2 eV must extend to higher 
energy.}
\label{fig4}
\end{figure}
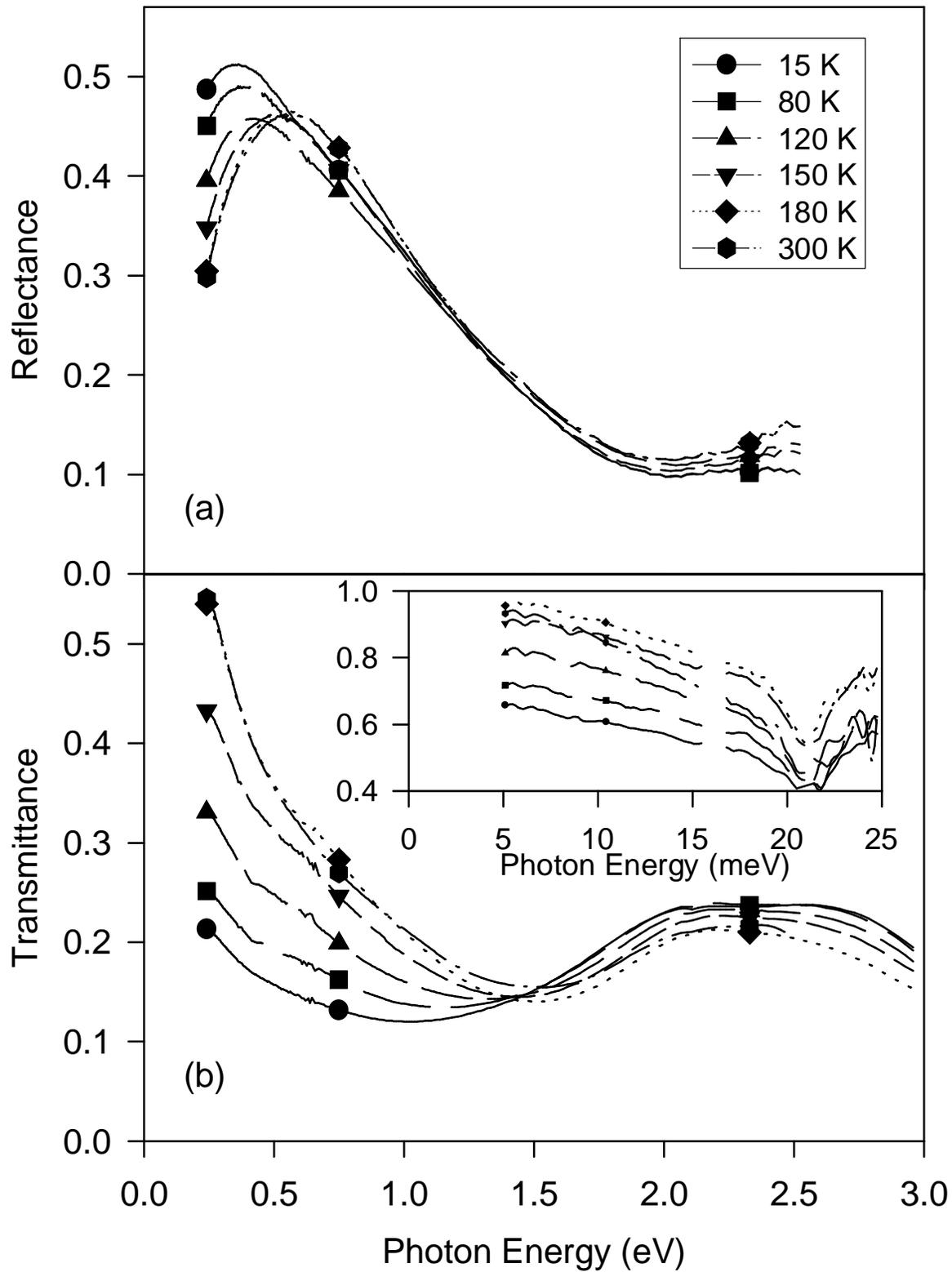
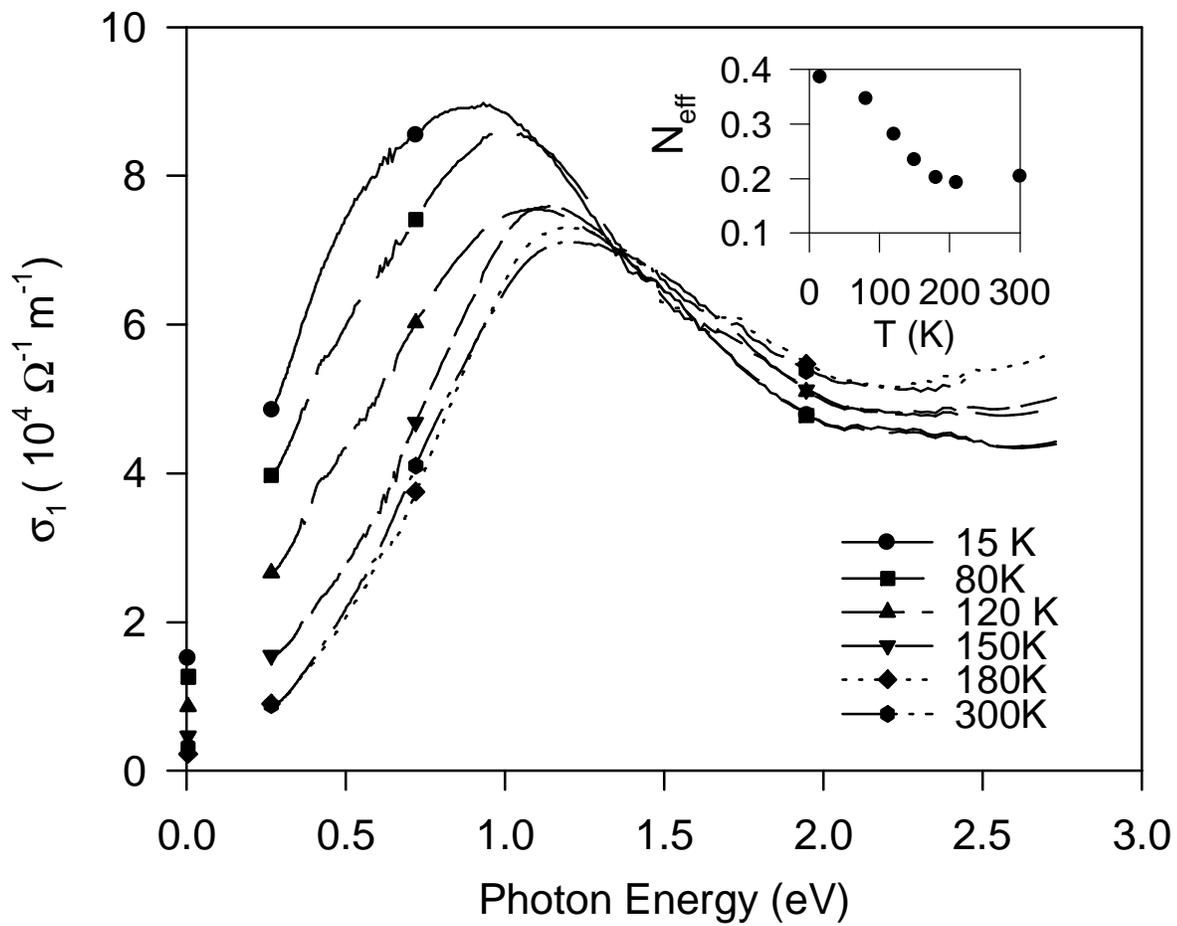
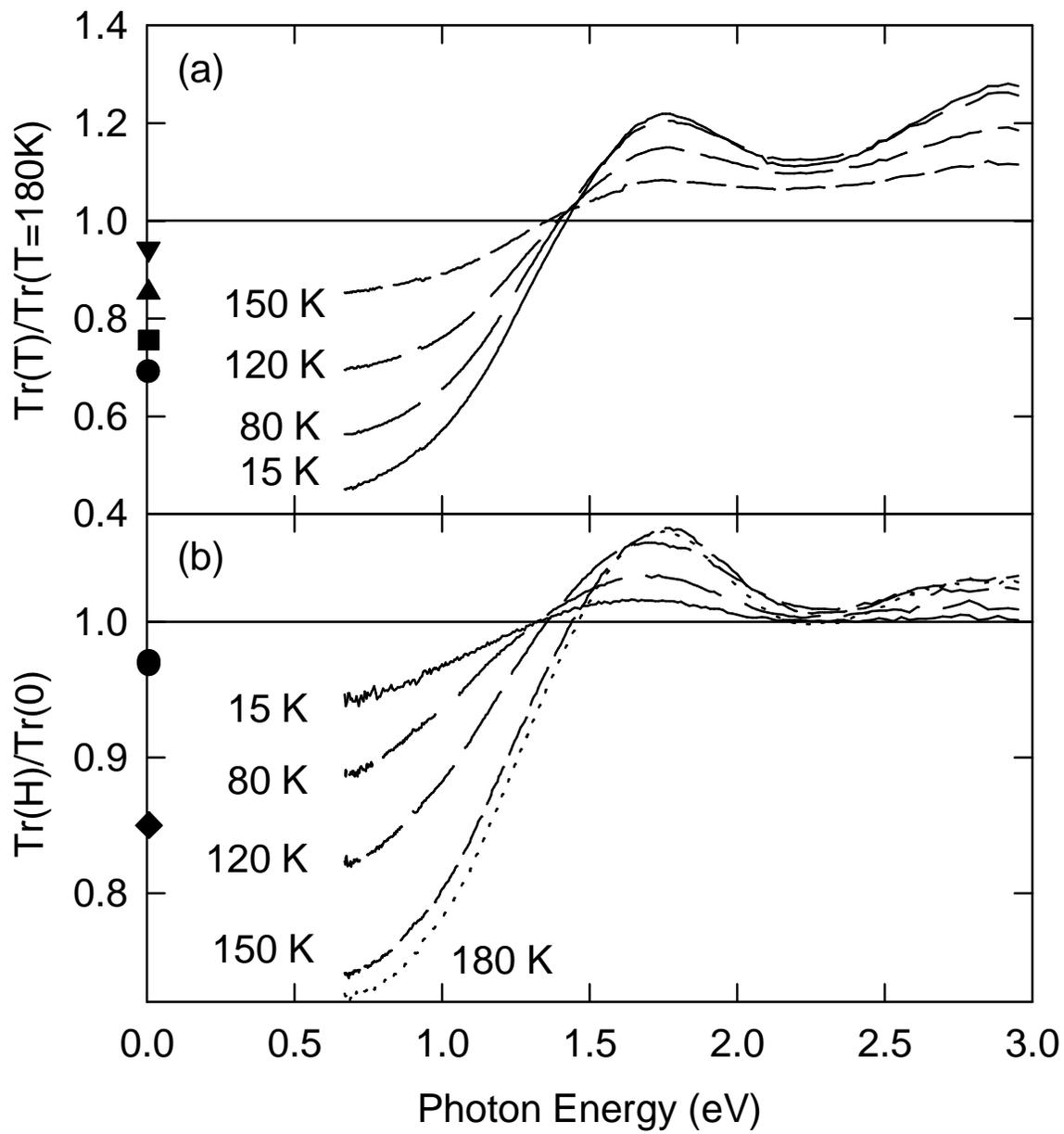
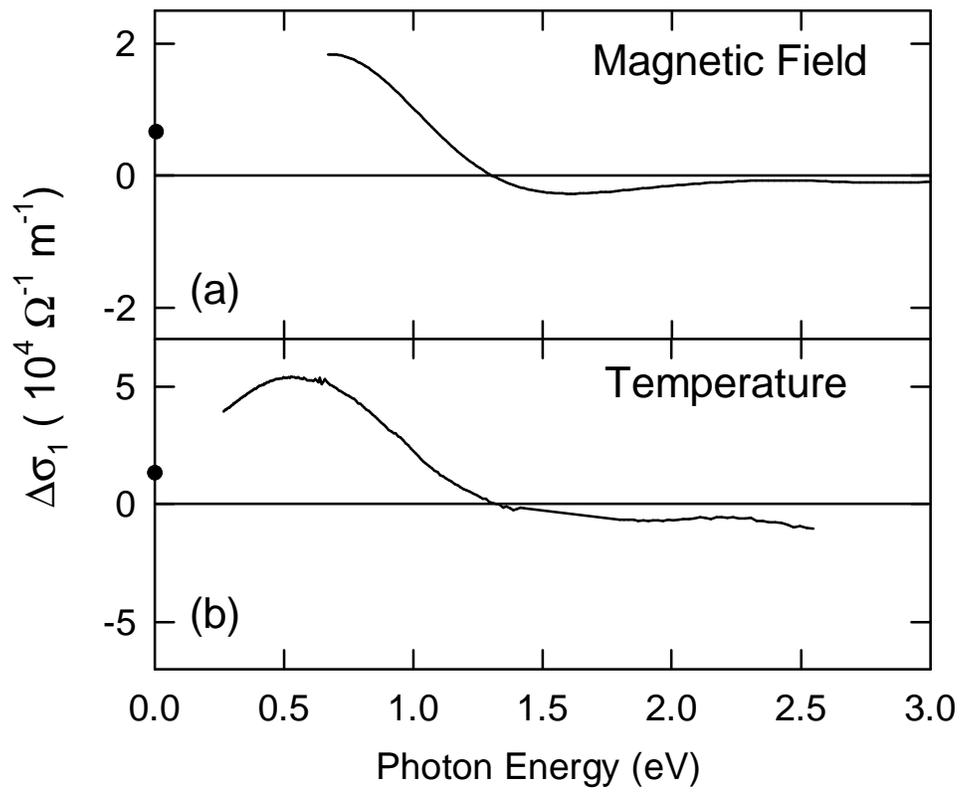
\end{document}